\documentclass[10pt]{iopart}
\usepackage{hyperref}
\usepackage{graphicx}
\usepackage{subfig}
\usepackage{color}
\usepackage{ulem}
\usepackage{amssymb}
\usepackage{cite}

\begin{document}

\title{Nonlinear saturation of reversed shear Alfv\'en eigenmode via high-frequency quasi-mode generation} 

\author{Zhiwen Cheng$^1$, Guangyu Wei$^1$, Lei Ye$^2$ and Zhiyong Qiu$^{2,3}$}
\address{$^1$ Institute for Fusion Theory and Simulation, School of Physics, Zhejiang University, Hangzhou, China}
\address{$^2$ CAS Key Laboratory of Frontier Physics in Controlled Nuclear Fusion and  Institutes of Plasma Physics, Chinese Academy of Sciences, Hefei 230031, People's Republic of China}
\address{$^3$ Center for Nonlinear Plasma Science and C.R. ENEA Frascati, C.P. 65, 00044 Frascati, Italy}
\ead{zqiu@ipp.ac.cn}

\begin{abstract}
    A    nonlinear saturation mechanism for reversed shear Alfv\'en eigenmode (RSAE)  is proposed and analysed, and is shown to be of relevance to typical reactor parameter region. The saturation is achieved through the generation of high-frequency quasi-mode due to nonlinear coupling of two RSAEs,  which is  then damped due to  coupling with  the shear Alfv\'en continuum, and leads to the nonlinear saturation of the primary RSAEs .   An estimation of the nonlinear damping rate is also provided. 
\end{abstract}
\noindent{reversed shear Alfv\'en eigenmode, nonlinear mode coupling, continuum damping, gyrokinetic theory}

\maketitle
\ioptwocol

\section{Introduction}
Energetic particles (EPs) including fusion alpha particles are of crucial importance   in magnetically confined fusion plasmas due to their contribution to  plasma heating and potentially 
current drive \cite{AFasoliNF2007,LChenRMP2016}. A key aspect of EP confinement is related to the shear Alfv\'en wave  (SAW) instabilities  \cite{HAlfvenNature1942} resonantly excited by EPs \cite{YKolesnichenkoVAE1967,AMikhailovskiiSPJ1975,MRosenbluthPRL1975,LChenPoP1994,LChenNF2007a}. 
In magnetic confinement devices, SAW can be excited as various EP continuum modes (EPMs) \cite{LChenPoP1994} or discrete Alfv\'en modes (AEs) \cite{CZChengAP1985,LChenVarenna1988,GFuPoFB1989} inside the frequency gaps of the SAW continuum induced by equilibrium magnetic geometry and plasma nonuniformity. 
These SAW instabilities can then induce significant EPs anomalous transport loss across the magnetic surfaces, leading to plasma performance degradation and even damage of plasma facing components \cite{IPBNF1999,RDingNF2015}. 
With the EPs transport rate determined by the saturation amplitude and spectrum of SAW instabilities \cite{LChenJGR1999,MFalessiPoP2019}, it is crucial to understand the nonlinear dynamics resulting in their saturation. 
In the past decades, nonlinear saturation of SAW instabilities has been broadly investigated both numerically  and theoretically \cite{HBerkPoFB1990c,YTodoPoP1995,JLangPoP2010,SBriguglioPoP2014,JZhuPoP2013,HZhangNF2022,DSpongPoP1994,TSHahmPRL1995,FZoncaPRL1995,LChenPPCF1998,YTodoNF2010,LChenPRL2012,ZQiuPoP2016,ZQiuNF2017,ZQiuPRL2018,ZQiuNF2019a,LChenNF2022,LChenNF2023}, 
among which,  one of the most important channel  is  nonlinear wave-wave coupling \cite{RSagdeevbook1969,ZQiuRMPP2023}, i.e. nonlinear spectrum evolution of SAW instabilities due to interacting with other collective  electromagnetic oscillations.

In the advanced scenarios of  future reactor burning plasmas, a large fraction of non-inductive (e.g. bootstrap) current will be maintained \cite{GormezanoNF2007} off-axis,  and the magnetic shear is reversed in the plasma core region, where large fraction of energetic fusion alpha particles are generated \cite{TWangPoP2018}.  As a result, a specific Alfv\'en eigenmode, namely the reversed shear Alfv\'en eigenmode (RSAE, also known as Alfv\'en cascade due to its frequency sweeping  character \cite{KShinoharaNF2001,SSharapovNF2006,HBerkPRL2001,FZoncaPOP2002}) could be excited and play important roles in transport of fusion alpha particles.  In particular, as multiple-$n$ RSAEs can be strongly driven unstable simultaneously in reactors with machine size being much larger than fusion alpha particle  characteristic orbit width,   RSAEs can lead to strong alpha particle re-distribution and transport \cite{TWangPoP2018,TWangPoP2019}.    RSAE is    a branch of Alfv\'en eigenmodes localized around  the  SAW continuum extremum induced by  the local  minimum of the safety factor $q$-profile (labeled as $q_{min}$) to minimize the continuum damping,  and is  characterized by a radial width of $\sim \sqrt{q/(r_0^2q^{''})}$ \cite{FZoncaPOP2002}, with $r_0$ being the radial location of $q_{min}$ and $q^{''}\equiv \partial^2_r q$. 
 RSAE was originally  observed in the advanced operation experiments in JT-60U tokamak  \cite{HKimuraNF1998}, and was then detected in numerous JET discharges \cite{SSharapovPLA2001}. In present day tokamaks, RSAEs are generally excited by large orbit EP during current ramp up stage where  reversed shear $q$-profile is created by insufficient current penetration \cite{JHuangNF2020}. 
In most cases, with $q_{min}$ decreasing from, e.g.,  a rational value $m/n$ to $(m-1/2)/n$, the RSAEs exhibit  upward frequency sweeping from beta-induced Alfv\'en eigenmode (BAE) \cite{FZoncaPPCF1996} to the toroidal Alfve\'n eigenmode (TAE) \cite{CZChengAP1985,GFuPoFB1989} frequency ranges. Here, $m$ and $n$ stand for the poloidal and toroidal mode numbers,  respectively. 

Due to the increasing importance in reactor burning plasmas operating at advanced scenarios, RSAE has drawn much  attention in recent investigations. 
For instance, the resonant decay of RSAE into a generic low frequency Alfv\'en mode (LFAM) was investigated in Ref. \citenum{SWeiNF2022}, based on which,  a potential alpha channelling mechanism \cite{NFischNF1994} via the LFAM Landau damping was also proposed and analysed. 
The modulational instability of a finite amplitude RSAE and excitation of the zero-frequency zonal structures were investigated in Ref. \citenum{SWeiJPP2021}, where RSAE was saturated due to the modulation of SAW continuum and scattering into short-radial-wavelength stable domain. In particular, it is pointed out that, the generation of zonal current around the $q_{min}$ region can be of particular importance, due to the sensitive dependence of RSAE to reversed shear profiles. 
Further numerical  investigations of RSAE nonlinear dynamics can also be found  in, e.g., Refs. \cite{TWangPST2024,PLiuRMPP2023}, where RSAE nonlinear saturation due to wave-particle radial decoupling and zonal flow generation were investigated, respectively. It is also noteworthy that, in Ref. \citenum{LYeNF2023}, a nonlinear saturation channel of RSAEs via nonlinear harmonic generation was investigated, where quasi-modes with double and/or triple toroidal mode numbers of the primary linearly unstable RSAE were generated due to kinetic electron contribution via ``magnetic fluttering", and led to RSAE nonlinear saturation.  The setting of the simulation seems  though, to some extend  ``artificial", as  only few toroidal mode numbers are kept in the simulation, it provides the important information of RSAE dissipation via nonlinear harmonic generation.

Motivated  by Ref. \citenum{LYeNF2023},  in this  work, we present a potential nonlinear saturation mechanism for RSAE via nonlinear quasi-mode generation. This nonlinear mode coupling is achieved through the non-adiabatic 
responses to electrons, corresponding to the magnetic fluttering nonlinearity as addressed in Ref. \citenum{LYeNF2023}. Meanwhile, the   mode coupling is generalized  from RSAE self-coupling in the simulation \cite{LYeNF2023} to include also the interaction between two RSAEs with  different toroidal mode numbers, for which the nonlinear coupling could be much stronger. 
 Generally, this quasi-mode could  experience  significant continuum or radiative  damping,  and provide a channel for primary RSAE energy dissipation. 
Using nonlinear gyrokinetic theory, the parametric dispersion relation of this nonlinear mode coupling process is derived. Focusing on the continuum damping of the quasi-mode, the resultant nonlinear damping to RSAE is then analyzed and estimated.

The remainder of this paper is arranged as follows. In Sec. \ref{sec:model}, the theoretical model is given. In Sec. \ref{sec:NL_eq}, the nonlinear  dispersion relation describing the RSAE nonlinear evolution due to interaction with another background RSAE is derived. 
It is then used in Sec. \ref{sec:ana} to investigate the continuum damping of the quasi-mode. An estimation of the resulting damping to RSAE is also presented.

\section{Theoretical Model}\label{sec:model}

\begin{figure*}
    \centering
    \begin{minipage}[]{.3\linewidth}
        \includegraphics[scale=0.25]{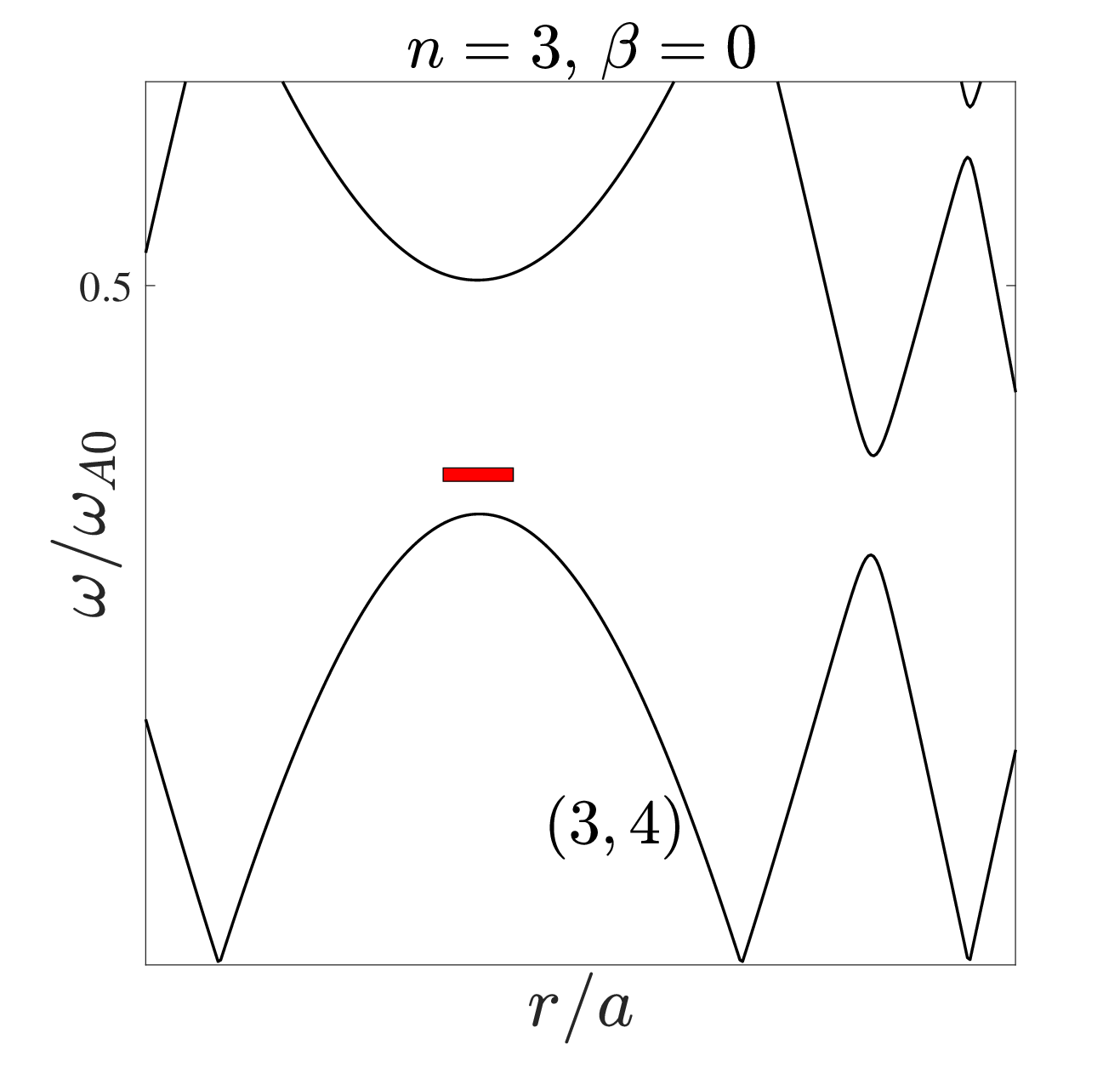}
    \end{minipage}
    \begin{minipage}[]{.3\linewidth}
        \includegraphics[scale=0.25]{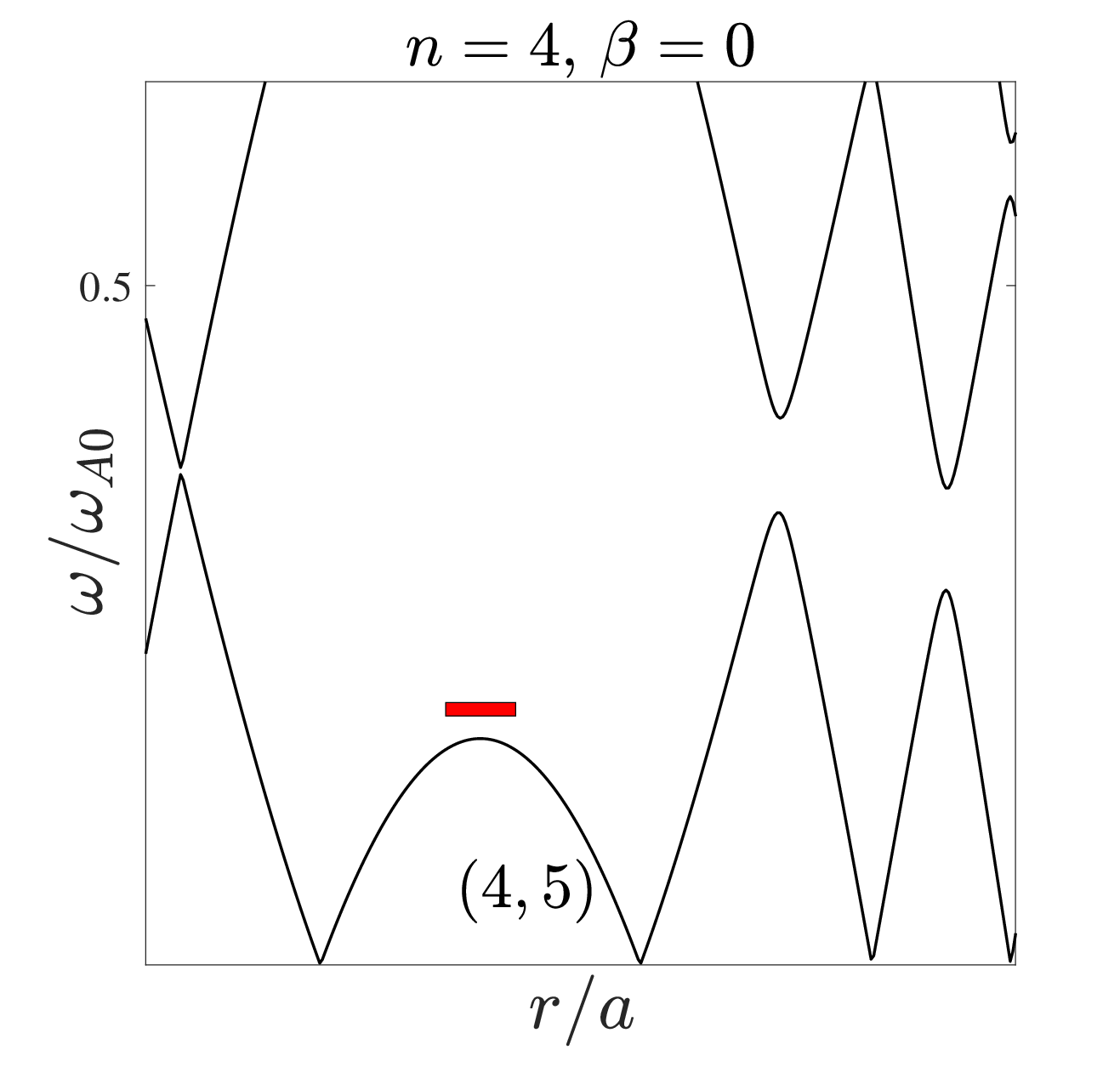}
    \end{minipage}
    \begin{minipage}[]{.3\linewidth}
        \includegraphics[scale=0.25]{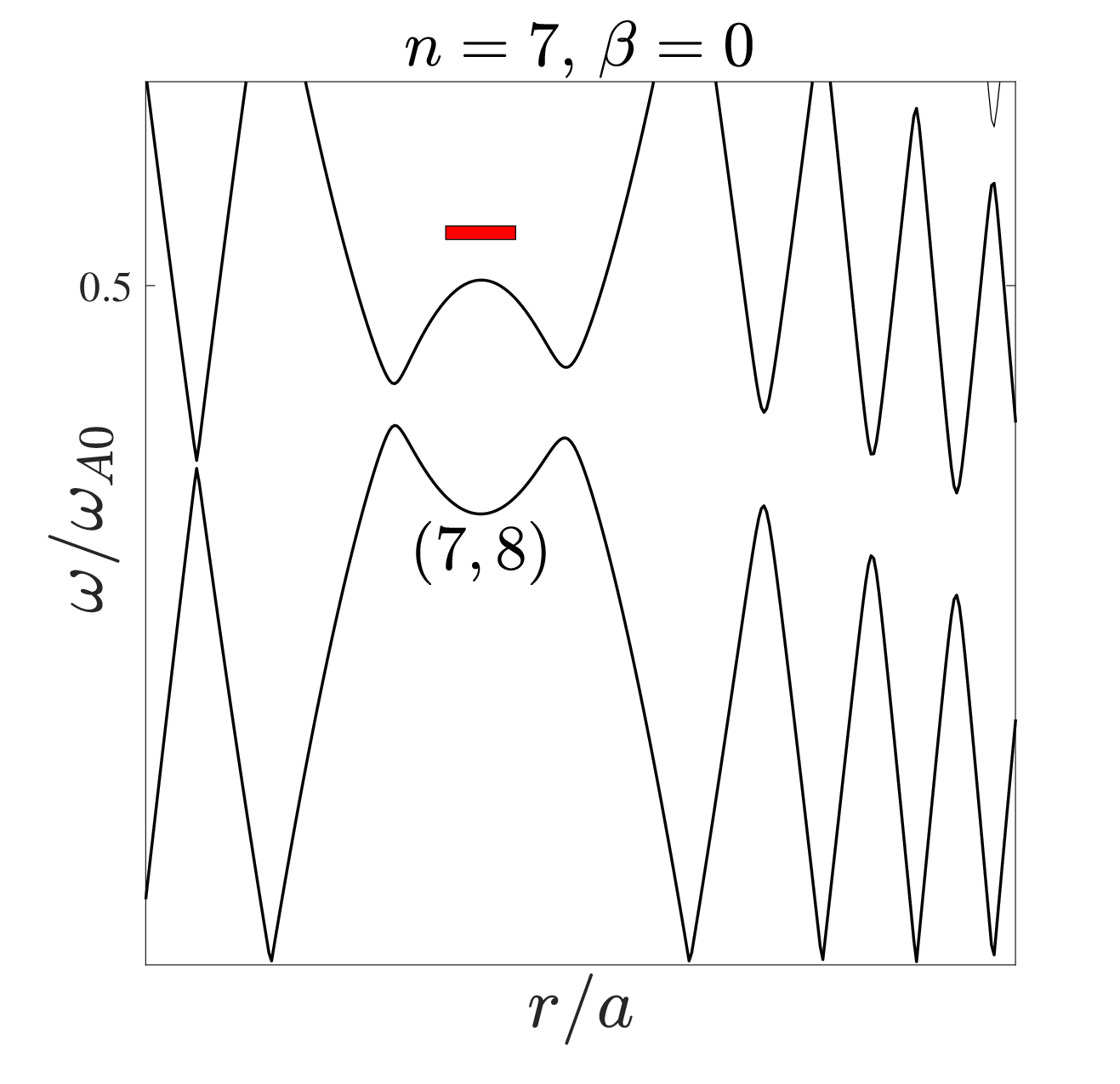}
    \end{minipage}
    \caption{Schematic plots of Alfv\'en continuum for $n=3,4$ and $7$ respectively. With the red bars stand for Alfv\'en modes, (a) and (b) represent RSAEs, and (c) represents the high-frequency quasi-mode generated by the nonlinear coupling of two RSAEs. Here, the horizontal axis is $r/a$ in arbitrary units, the vertical axis is the normalized frequency to $\omega_{A0}$, i.e. the Alfv\'en frequency on the major axis.}
    \label{fig:continuum}
\end{figure*}

Considering two co-propagating  RSAEs  $\Omega_0\equiv\Omega_0(\omega_0,\mathbf{k}_1)$ and $\Omega_1\equiv\Omega_1(\omega_1,\mathbf{k}_1)$ coupling and generating a beat wave $\Omega_b\equiv\Omega_b(\omega_b,\mathbf{k}_b)$, 
with the frequency and wavenumber of $\Omega_b$ determined by the matching condition $\Omega_b=\Omega_0 +\Omega_1$,  the beat wave $\Omega_b$ is   likely a high-frequency quasi-mode bearing significant continuum  or 
radiative damping,  as it may not satisfy the global RSAE dispersion relation with  corresponding toroidal mode number $n_b=n_0+n_1$. Here, for ``co-propagating", we mean the two RSAEs propagate in the same direction along the magnetic field line.  This nonlinear coupling provides the primary   RSAEs an indirect damping mechanism,  and may result in their saturation. Here, for simplicity of the discussion while focusing on the main physics picture, we focus on the 
continuum damping of the quasi-mode,  and investigate the resultant nonlinear saturation of primary RSAEs. A sketched  illustration of the proposed process   is given in Fig. \ref{fig:continuum}, where two RSAEs with $n=3,4$  couple  and generate an  $n=7$ high-frequency quasi-mode, which can be heavily damped due to coupling  with the corresponding shear Alfv\'en continuum.  

The nonlinear coupling of the two RSAEs and the resultant damping are investigated in a uniform low-$\beta$ magnetized plasma using the standard nonlinear perturbation theory, with $\beta\ll1$ being the ratio between plasma and magnetic pressures.  Here, for ``uniform", we mean the effects associated with diamagnetic effects are systematically neglected, while noting magnetically confined plasma is intrinsically nonuniform. 
Introducing the scalar potential $\delta\phi$ and parallel component of vector potential $\delta A_\parallel$ as the perturbed field variables, one then has $\delta\phi=\delta\phi_0+\delta\phi_1+\delta\phi_b$ 
with the subscripts $0$, $1$ and $b$ denoting $\Omega_0$, $\Omega_1$ and $\Omega_b$, respectively. For convenience of investigation, $\delta A_\parallel$ is replaced by $\delta\psi\equiv \omega\delta A_\parallel/(ck_\parallel)$, such that $\delta\phi=\delta\psi$ can straightforwardly recover 
the ideal MHD limit, i.e. vanishing parallel electric field fluctuation $\delta E_\parallel$. Both electrons and ions are chacterised by  Maxwellian equilibrium distributions $F_M$.

For RSAEs typically dominated by single-$n$ and single-$m$ mode structures near $q_{min}$, we take
\begin{equation*}
    \delta\phi_k =A_k(t) \Phi_k(x) \exp(-i\omega_kt+in\xi-im\theta),
\end{equation*}
with $A_k(t)$ being the slowly varying mode amplitude, $\Phi_k(x)$ being the parallel mode structure localized around $q_{min}$ with $x\equiv nq-m$, and 
the normalization   $\int |\Phi_k|^2dx=1$ can be adopted.

Nonlinear mode  equations can be derived from charge  quasi-neutrality condition
\begin{equation}
    \frac{N_0e^2}{T_i}\left(1+\frac{T_i}{T_e}\right)\delta\phi_k=\sum_{j=e,i}\langle qJ_k\delta H_k\rangle_j,
\end{equation}
and   nonlinear gyrokinetic vorticity equation \cite{LChenJGR1991}
\begin{eqnarray}
   && \frac{c^2}{4\pi\omega_k^2}B\frac{\partial}{\partial l}\frac{k^2_\bot}{B}\frac{\partial}{\partial l}\delta\psi_k+\frac{e^2}{T_i}
   \langle(1-J_k^2)F_M\rangle\delta\phi_k \nonumber\\
    &&-\sum_{j=e,i}\left \langle qJ_k\frac{\omega_d}{\omega_k}\delta H_k\right \rangle_j \nonumber\\
    &=&-\frac{i}{\omega_k}\sum_{\mathbf{k=k'+k''}}\Lambda_{k''}^{k'}[\langle e(J_k J_{k'}-J_{k''})\delta L_{k'}\delta H_{k''}\rangle \nonumber \\
   && +\frac{c^2}{4\pi}k^{''2}_\bot\frac{\partial_l\delta\psi_{k'}\partial_l\delta\psi_{k''}}{\omega_{k'}\omega_{k''}}].\label{eq:vorticity}
\end{eqnarray}
Here, the terms on the left hand side  of  equation (\ref{eq:vorticity}) are field line bending, inertia and curvature-pressure coupling terms, respectively, whereas the terms on the right hand side represent Reynolds and Maxwell stresses   dominating in short wavelength limit. 
$N_0$ is the equilibrium particle density, $q_j$ is the electric charge, the angular brackets $\langle \dots\rangle$ denote velocity space integration, $\partial_l$ is the spatial derivative along the equilibrium magnetic field, $k_\bot=\sqrt{k^2_r+k^2_{\theta}}$ is  the perpendicular wavenumber, $J_k\equiv J_0(k_\bot \rho)$ with $J_0$ being the Bessel function of zero index accounting for finite Larmor radius effects, and $\rho=v_\bot/\Omega_c$ is the Larmor radius with $\Omega_c$ being the cyclotron frequency.  Furthermore, 
$\omega_d=(v^2_\bot+2v^2_\parallel)/(2\Omega_cR_0)(k_r \sin\theta+k_\theta\cos\theta)$ is  the magnetic drift frequency,  $\Lambda_{k''}^{k'}=(c/B_0)\hat{\mathbf{b}}\cdot\mathbf{ k''\times k'}$ accounts for perpendicular coupling with the constraint of frequency and wavevector matching conditions, 
and $\delta L_k\equiv\delta \phi_k-k_\parallel v_\parallel\delta\psi_k/\omega_k$ is the scalar potential in the frame moving with guiding center.  
The non-adiabatic particle response  $\delta H_k$ is  derived from the  nonlinear gyrokinetic equation \cite{EFriemanPoF1982}:
\begin{eqnarray}
    (-i\omega+v_\parallel\partial_l+i\omega_d)\delta H_{k}=-i\frac{q}{T_j}\omega_kF_MJ_k\delta L_k \nonumber\\
    -\sum_{\mathbf{k=k'+k''}}\Lambda_{k''}^{k'}J_{k'}\delta L_{k'}\delta H_{k''}.\label{eq:GKE}
\end{eqnarray}

\section{Nonlinear mode  equations}\label{sec:NL_eq}
In this section, the coupled  nonlinear   equations for RSAE and the high-frequency quasi-mode are derived in Secs. \ref{sec:Omegab} and \ref{sec:Omega0}, which are then combined and give the parametric dispersion relation in Sec. \ref{sec:PDI_DR}.

\subsection{Nonlinear quasi-mode  $\Omega_b$ genaration}\label{sec:Omegab}
The nonlinear   equation for  quasi-mode $\Omega_b$  generation can be derived from the quasi-neutrality condition and the nonlinear gyrokinetic vorticity equation. 
The nonlinear non-adiabatic particle response of $\Omega_b$ can be derived from the nonlinear component of equation (\ref{eq:GKE}) noting the  $k_\parallel v_e\gg\omega\gg k_\parallel v_i\gtrsim\omega_d$ ordering, and one obtains 
\begin{eqnarray}
    \delta H_{bi}^{NL}&=&0, \label{eq:biNL}\\
    \delta H_{be}^{NL}&=&i\Lambda_0^1 \frac{e}{T_e}F_M  \frac{1}{k_{\parallel b}}\left(\frac{k_{\parallel 1}}{\omega_1}-\frac{k_{\parallel 0}}{\omega_0}\right)\delta\psi_0\delta\psi_1.
    \label{eq:beNL}
\end{eqnarray}
In deriving equations (\ref{eq:biNL}) and (\ref{eq:beNL}), the linear particle responses  $\delta H_{ki}^L=(e/T_i)F_{M}J_k\delta\phi_k$ and $\delta H_{ke}^L=-(e/T_e)F_{M}\delta\psi_k$ are used. 
 With $\delta H_{be}^{NL}$ representing the coupling between $\Omega_0$ and $\Omega_1$ due to nonlinear electron contribution, it corresponds to the magnetic fluttering nonlinearity investigated  in \cite{LYeNF2023}. 
Substituting equations (\ref{eq:biNL}) and (\ref{eq:beNL}) into the quasi-neutrality condition, one obtains
\begin{equation}
    \delta\psi_b=\delta\phi_b+i\Lambda_0^1 \frac{1}{k_{\parallel b}}\left(\frac{k_{\parallel 1}}{\omega_1}-\frac{k_{\parallel 0}}{\omega_0}\right)\delta\psi_0\delta\psi_1,\label{eq:QN_b}
\end{equation}
i.e. breaking of ideal MHD constraint due to nonlinear mode coupling, while finite parallel electric field associated with linear FLR effects is not included here for simplicity \cite{AHasegawaPoF1976}. This is also consistent with the $b_k\ll1$ ordering for linear unstable RSAEs with typically $k^{-1}_{\perp}$ comparable to EP characteristic orbit width. 
Substituting the particle responses into the nonlinear gyrokinetic vorticity equation, one obtains
\begin{eqnarray}
    &&b_b\left(\delta\phi_b-\frac{k_{\parallel b}^2v_A^2}{\omega_b^2}\delta\psi_b-\frac{\omega_G^2}{\omega_b^2}\delta\phi_b\right) \nonumber \\ 
    &=&-i\frac{\Lambda_0^1}{\omega_b}(b_0-b_1)\left(1-\frac{k_{\parallel 0}k_{\parallel 1}v_A^2}{\omega_0\omega_1}\right)\delta\phi_0\delta\phi_1,\label{eq:GV_b}
\end{eqnarray}
with $b_k=k_\bot^2\rho_i^2/2$, $v_A$ being the Alfv\'en speed, $\omega_G\equiv \sqrt{7/4+\tau}v_i/R_0$ being the leading order geodesic acoustic mode frequency \cite{NWinsorPoF1968,FZoncaEPL2008} and $\tau\equiv T_e/T_i$.
Combining equations (\ref{eq:QN_b}) and   (\ref{eq:GV_b}), one obtains 
\begin{equation}
    b_b \varepsilon_{Ab}\delta\phi_b=i\frac{\Lambda_0^1}{\omega_b}\beta_b\delta\phi_0\delta\phi_1.
    \label{eq:NLcpb}
\end{equation} 
Equation (\ref{eq:NLcpb}) is the desired nonlinear   equation describing the high-frequency quasi-mode  $\Omega_b$ generation  due to $\Omega_0$ and $\Omega_1$ coupling,
with the $\Omega_b$ dielectric function  $\varepsilon_{Ab}$ defined as
\begin{eqnarray*}
    \varepsilon_{Ab}\equiv 1-\frac{k_{\parallel b}^2v_A^2}{\omega_b^2}-\frac{\omega_G^2}{\omega_b^2},
\end{eqnarray*}
and the nonlinear coupling coefficient $\beta_b$ given by
\begin{eqnarray*}
    \beta_b=b_b \frac{k_{\parallel b}v_A}{\omega_b}\left(\frac{k_{\parallel 1}v_A}{\omega_1}-\frac{k_{\parallel 0}v_A}{\omega_0}\right) \nonumber \\
    -(b_0-b_1)\left(1-\frac{k_{\parallel 0}k_{\parallel 1}v^2_A}{\omega_0\omega_1}\right).
\end{eqnarray*}
It is worth mentioning that, the $\Omega_b$ dielectric function,  $\varepsilon_{Ab}$, may not satisfy the global linear RSAE dispersion relation for toroidal mode number $n_b$, 
 and $\Omega_b$ could  be a quasi-mode experience  heavy   damping, leading to the dissipation  of   both itself  and the primary  RSAEs, as shown later.

\subsection{Nonlinear equation for $\Omega_0$}\label{sec:Omega0}

The nonlinear coupling equation for the test RSAE $\Omega_0$ can be derived following a similar procedure. However, noting that $\Omega_b$ is a quasi-mode, one needs to keep both the linear and nonlinear particle responses since they could  be of the same order. 
The resultant nonlinear non-adiabatic particle responses of $\Omega_0$ are respectively
\begin{eqnarray}
\delta H_{0i}^{NL}&=&0,
    \label{eq:0iNL}\\
    \delta H_{0e}^{NL}&=&-i\Lambda_0^1  \frac{e}{T_e}F_M \frac{1}{k_{\parallel 0}}\left(\frac{k_{\parallel 1}}{\omega_1}-\frac{k_{\parallel b}}{\omega_b}\right)\delta\psi_1^{*}\delta\psi_b \nonumber \\ 
 &-&(\Lambda_0^1)^2\frac{e}{T_e}F_M \frac{k_{\parallel 1}}{k_{\parallel 0}k_{\parallel b}\omega_1}\left(\frac{k_{\parallel 1}}{\omega_1}-\frac{k_{\parallel 0}}{\omega_0}\right)|\delta\psi_1|^2\delta\psi_0.
    \label{eq:0eNL}
\end{eqnarray}
Substituting equations (\ref{eq:0iNL}) and (\ref{eq:0eNL}) into the quasi-neutrality condition, one obtains
\begin{eqnarray}
    \delta\psi_0&=&\delta\phi_0-i\Lambda_0^1 \frac{1}{k_{\parallel 0}}\left(\frac{k_{\parallel 1}}{\omega_1}-\frac{k_{\parallel b}}{\omega_b}\right)\delta\psi_1^{*}\delta\psi_b  \nonumber \\ 
    &-&(\Lambda_0^1)^2\frac{k_{\parallel 1}}{k_{\parallel 0}k_{\parallel b}\omega_1}\left(\frac{k_{\parallel 1}}{\omega_1}-\frac{k_{\parallel 0}}{\omega_0}\right)|\delta\psi_1|^2\delta\psi_0.\label{eq:QN_0}
\end{eqnarray}
On the other hand,   the nonlinear gyrokinetic vorticity equation yields 
\begin{eqnarray}
    &&b_0\left(\delta\phi_0-\frac{k_{\parallel 0}^2v_A^2}{\omega_0^2}\delta\psi_0-\frac{\omega_G^2}{\omega_0^2}\delta\phi_0\right) \nonumber \\
    &=&i\frac{\Lambda_0^1}{\omega_0}(b_b-b_1)\left(1-\frac{k_{\parallel b}k_{\parallel 1}v_A^2}{\omega_b\omega_1}\right)\delta\phi_1^{*}\delta\phi_b.\label{eq:GV_0}
\end{eqnarray}
Combining equations (\ref{eq:QN_0}) and   (\ref{eq:GV_0}), one obtains the nonlinear equation of $\Omega_0$
\begin{equation}
    b_0 (\varepsilon_{A0}+\varepsilon_{A0}^{NL})\delta\phi_0=-i\frac{\Lambda_0^1}{\omega_0}\beta_0\delta\phi_1^{*}\delta\phi_b,
    \label{eq:NLcp0}
\end{equation}
with the linear $\Omega_0$ dielectric function in the WKB limit given by 
\begin{eqnarray}
\varepsilon_{A0}\equiv 1-\frac{k_{\parallel 0}^2v_A^2}{\omega_0^2}-\frac{\omega_G^2}{\omega_0^2},\nonumber
\end{eqnarray}
the nonlinear coupling coefficient $\beta_b$ given by
\begin{eqnarray}
    \beta_b=b_0 \frac{k_{\parallel 0}v_A}{\omega_0}\left(\frac{k_{\parallel 1}v_A}{\omega_1}-\frac{k_{\parallel b}v_A}{\omega_b}\right) \\
    -(b_b-b_1)\left(1-\frac{k_{\parallel b}k_{\parallel 1}v^2_A}{\omega_b\omega_1}\right),\nonumber
\end{eqnarray}
and  $\varepsilon_{A0}^{NL}$ due to nonlinear particle contribution to $\Omega_b$ being
\begin{eqnarray}
    \varepsilon_{A0}^{NL}=\frac{(\Lambda_0^1)^2}{\omega_0\omega_b}\frac{k_{\parallel0}k_{\parallel1}v^2_A}{\omega_0\omega_1}\frac{\omega_b}{k_{\parallel b}}\left(\frac{k_{\parallel 1}}{\omega_1}-\frac{k_{\parallel 0}}{\omega_0}\right)|\delta\phi_1|^2.\nonumber
\end{eqnarray}
Equation (\ref{eq:NLcp0}) is the  nonlinear   equation for the test RSAE $\Omega_0$ evolution due to  the feedback of the quasi-mode $\Omega_b$,  and can be coupled with   equation (\ref{eq:NLcpb}) to yield the nonlinear dispersion relation for $\Omega_0$ regulation via the high-frequency quasi-mode generation. 

\subsection{Parametric dispersion relation}\label{sec:PDI_DR}
Combining equations (\ref{eq:NLcpb}) and   (\ref{eq:NLcp0}),   one obtains
\begin{equation}
    b_0 b_b (\varepsilon_{A0}+\varepsilon_{A0}^{NL})\varepsilon_{Ab}\delta \phi_0 = \frac{(\Lambda_0^1)^2}{\omega_0 \omega_b}\beta_0 \beta_b|\delta\phi_1|^2\delta\phi_0.
    \label{eq:PDR}
\end{equation}
Equation (\ref{eq:PDR}) describes  the evolution of the test RSAE $\Omega_0$ due to the nonlinear interaction with another RSAE $\Omega_1$, which can also be considered as the ``parametric decay dispersion relation" of $\Omega_1$ decaying into $\Omega_0$ and $\Omega_b$. 
Noting that  $\varepsilon_{A0}^{NL}$ related term contributes only  to the nonlinear frequency shift, re-organising equation (\ref{eq:PDR}) and taking the imaginary part, one obtains
\begin{equation}
    \frac{2\gamma_{AD}}{\omega_{0r}}b_0 \delta\phi_0=\pi\frac{(\Lambda_0^1)^2}{\omega_0 \omega_b}\frac{\beta_0 \beta_b}{b_b}\delta(\varepsilon_{Ab})|\delta\phi_1|^2\delta\phi_0. 
    \label{eq:gamma}
\end{equation}
In deriving equation (\ref{eq:gamma}), we have expanded $\varepsilon_{A0}\simeq i\partial_{\omega_{0r}}\varepsilon_{A0}(\partial_t-\gamma_{0})\simeq -(2i/\omega_{0r})\gamma_{AD}$,  with $\gamma_0$  being the  linear growth rate of $\Omega_0$ and   $\gamma_{AD}$ being its  damping rate   due to scattering by $\Omega_b$,  respectively.  It is also noteworthy that, as $\Omega_b$ is a quasi-mode with the imaginary  part of $\varepsilon_{Ab}$ being comparable to the real part,  no expansion is made to $\varepsilon_{Ab}$.  
Meanwhile, for the continuum damping of interest, $Im (1/\varepsilon_{Ab})=-\pi\delta(\varepsilon_{Ab})$ is taken, corresponding to the absorption of the nonlinear generated quasi-mode $\Omega_b$ near the SAW continuum resonance layer \cite{AHasegawaPoF1976}.

\section{Nonlinear damping to test RSAE}\label{sec:ana}
An estimation of the nonlinear damping rate is made to quantify the contribution of this nonlinear process. 
Multiplying  equation (\ref{eq:gamma}) with $\Phi_0^{*}$ and averaging over radial mode structure,    one obtains
\begin{eqnarray}
    \frac{2\gamma_{AD}}{\omega_{0r}}\langle \Phi_0^{*}b_0\Phi_0\rangle_x \nonumber \\
    =\pi\left\langle \frac{(\Lambda_0^1)^2}{\omega_0 \omega_b}\frac{\beta_0 \beta_b}{b_b}\delta(\varepsilon_{Ab})|A_1|^2|\Phi_1|^2|\Phi_0|^2 \right\rangle_x. 
    \label{eq:gamma_int}
\end{eqnarray}
Here, $\langle \dots\rangle_x\equiv \int\cdots dx$ denotes the integration over $x$, with the weighting of $|\Phi_0|^2$. To make analytical progress, the parallel mode structures for RSAEs are taken as $\Phi_k\simeq \exp(-x^2/2\Delta_k^2)/(\pi^{1/4}\Delta_k^{1/2})$ with $\Delta_k$ being the characteristic radial width of the parallel mode structures and one typically has  $\Delta_0\sim \Delta_1\lesssim\mathcal{O}(1)$. 

Equation (\ref{eq:gamma_int}) gives the test RSAE $\Omega_0$ damping rate due to coupling to a ``background" RSAE $\Omega_1$. 
As multiple-n RSAEs could be driven unstable simultaneously \cite{TWangPoP2018} at the same location,  all the background RSAEs interacting with $\Omega_0$ should be taken into account. 
Summation over all the RSAEs within strong or moderate coupling range  to the test RSAE $\Omega_0$,  and assuming  the integrated electromagnetic fluctuation  amplitude induced by  RSAEs   is of the same order as the background RSAEs,   the nonlinear damping rate can be estimated as 
\begin{equation}
    \frac{\gamma_{AD}}{\omega_{0r}}\sim \frac{b}{\Delta_0\Delta_1}  \left(\frac{qR_0}{\rho_i}\right)^2 \left|\frac{\delta B_r}{B_0}\right|^2 \frac{x_0^3}{\varpi^6}  
    \sim \mathcal{O}(10^{-3}-10^{-2}).
    \label{eq:gamma_order}
\end{equation}
In estimating $\gamma_{AD}$, $\delta(\varepsilon_{Ab})=\partial\varepsilon_{Ab}/\partial x\sum_{x_0}^{}\delta(x-x_0)\simeq -2x/\varpi_b^2\sum_{x_0}^{}\delta(x-x_0)$ is taken, with $\varpi\equiv \omega/\omega_A$, $\omega_A\equiv v_A/(q_{min}R_0)$ being the local Alfv\'en frequency and $x_0$ being the zero points of $\varepsilon_{Ab}$.  
Other parameters are taken as $T_i/T_E\sim \mathcal{O}(10^{-2})$, $R_0/\rho_i\sim \mathcal{O}(10^3)$, $|\delta B_r/B_0|^2\sim\mathcal{O}(10^{-7})$, $b\sim k^2_\theta\rho_i^2\sim(T_i/T_E)/q^2$ for linearly unstable RSAEs with $T_E$ being the EP characteristic energy. 

Equation (\ref{eq:gamma_order}) shows an appreciable nonlinear damping   to the test RSAE $\Omega_0$, which could make significant contribution to its nonlinear saturation. Note that in the present work, only the scattering to    high-frequency quasi-mode is taken into account. Nevertheless, one can generalise the analysis to include other  damping channels including other  nonlinear mode coupling mechanism \cite{SWeiNF2022,SWeiJPP2021,YWangPST2022}. This is, however, beyond the scope of the present work, focusing on providing an interpretation  to the simulation of Ref. \citenum{LYeNF2023}, with the generalisation to  include  coupling to background RSAEs with different toroidal mode numbers.

\section{Conclusion}
Motivated by recent simulation study \cite{LYeNF2023},  a novel mechanism for RSAE nonlinear saturation is proposed and analysed, which is achieved through generation of  a high-frequency quasi-mode  by the nonlinear mode coupling of two RSAEs. 
This high-frequency quasi-mode can be  significantly damped due to coupling to the corresponding SAW continuum, thus leads to  a nonlinear damping effect  to the RSAEs,  and promotes their nonlinear saturation. 
The nonlinear dispersion relation describing this nonlinear coupling process is  derived based on the nonlinear gyrokinetic theory. 
To estimate the relevance  of this nonlinear saturation mechanism to RSAE, an estimation of the nonlinear damping rate to the test RSAE is given by $\gamma_{AD}/\omega_{0r}\sim \mathcal{O}(10^{-3}-10^{-2})$ under typical parameters of the future burning plasmas. 
This result could be comparable with the typical RSAE linear growth rate excited by resonant EPs, and thus, demonstrate the significance of the nonlinear saturation mechanism  proposed here. 

The nonlinear coupling coefficient derived in this work is complicated, and depends on various conditions including the frequency, wavenumber, radial mode structure of the RSAEs and the structure of Alfv\'en continuum corresponding to the mode number of quasi-mode. This study seeks to estimate the relevance  of the nonlinear saturation mechanism and has not done a thorough investigation on the optimised parameter regimes for this process to occur and dominate. For more detailed analysis, interested readers may refer to \cite{SWeiNF2022} with the nonlinear coupling coefficient having similar features.

As a final remark, the high-frequency quasi-mode discussed here, is damped due to the coupling to local Alfv\'en continuum only, whereas other damping effects (e.g. radiative damping and Landau damping due to frequency mismatch) are not included, and inclusion of which could yield an enhanced regulation effect to RSAE. 
Besides, the final nonlinear saturation of RSAE may require other channels including the self-consistent evolution of EPs distribution function \cite{TWangNF2020}, spontaneous decay into LFAM \cite{SWeiNF2022}, zonal field generation \cite{SWeiJPP2021} and geodesic acoustic mode (GAM) generation \cite{YWangPST2022}. 
Further comprehensive and detailed investigations, particularly through large scale nonlinear gyrokinetic simulations, are required to assess the saturation level of RSAE and the energetic particle transport rate.

\section*{Acknowledgement}
This work was  supported by  the National Science Foundation of China under Grant Nos. 12275236 and 12261131622, and  Italian Ministry for Foreign Affairs and International Cooperation Project under Grant  No. CN23GR02.

\section*{References}
\bibliographystyle{iopart-num}

\providecommand{\newblock}{}

\end{document}